\documentclass[11pt]{article}

\usepackage{graphicx}
\usepackage{latexsym}
\usepackage{amsfonts}
\usepackage{amsmath}
\usepackage{amssymb}

\newtheorem{theorem}{Theorem}[section]
\newtheorem{lemma}[theorem]{Lemma}
\newtheorem{observation}[theorem]{Observation}

\newcommand{\sq}{\hbox{\rlap{$\sqcap$}$\sqcup$}}
\newcommand{\qed}{\hspace*{\fill}\sq}

\newcommand{\ignore}[1]{ }

\newcommand{\algocomment}[1]{{\sffamily #1} }

\def\e{\ensuremath{\epsilon}}
\def\R{\ensuremath{\mathcal{R}}}
\def\eapprox{{\sffamily $\epsilon$-approx()}}
\def\esapprox{{\sffamily $\epsilon$-stream\_approx()}}
\def\weapprox{{\sffamily weighted\_$\epsilon$-approx()}}

\begin{document}

%=======================================================================
%  Title
%=======================================================================

\begin{center}
{\huge Deterministic Sampling and Range Counting in} \\[4pt]
{\huge Geometric Data Streams} \\[8pt]
{\large  Amitabha Bagchi,
  Amitabh Chaudhary,
  David Eppstein, and
  Michael T. Goodrich} \\[3pt]
Dept.~of Computer Science,
University of California,
Irvine, CA 92697-3425\ \ USA\\
{\tt \{bagchi,amic,eppstein,goodrich\}@ics.uci.edu}
\end{center}

%=====================================================================
%  Abstract
%=====================================================================

\begin{abstract}
We present memory-efficient deterministic algorithms
for constructing \e-nets and \e-approx\-i\-ma\-tions of 
streams of geometric data.
Unlike probabilistic approaches, these deterministic samples
provide guaranteed bounds on their approximation factors.  
We show how our deterministic samples can be used to answer approximate
online iceberg geometric queries on data streams.
We use these techniques to approximate several
robust statistics of geometric data streams, including Tukey depth, simplicial
depth, regression depth, the Thiel-Sen estimator, and the least
median of squares.
Our algorithms use only a polylogarithmic
amount of memory, provided the desired approximation
factors are inverse-polylogarithmic.
We also include a lower bound for non-iceberg geometric queries.
\end{abstract}

%=====================================================================
%  Introduction
%=====================================================================

\section{Introduction} 
\label{sec:introduction}

With the proliferation of streams of packets on the Internet, as well
as data streaming from embedded systems, digital monitors, sensor
networks, and scientific instruments, there is a need for new
algorithms that can compute approximations or answer approximate
queries on data streams.  The main challenge in these contexts is that
the data volumes are often much larger than the memory size of a
typical computer.  Thus, there is a considerable amount of interest in
methods that can process data streams using limited memory (e.g., see
recent surveys by Muthukrishnan~\cite{muthu-survey-03} and
Babcock~\cite{babcock-models_streams-02}).  The model we choose to
work in is the so called {\em Time Series} model in which each time
instant reveals a new element of the data stream ``signal.''

A typical approach in data streaming algorithms is to maintain a
random sample of the input data and perform computations on the sample
with the hope that information about the sample can be used to infer
properties of the entire set.  Naturally, such inferences come with an
associated probability that they are inaccurate.  In this paper, we
are interested in deterministically constructing samples of a data
stream that have guaranteed approximation properties for the original
set.  Moreover, because of the limited memory restriction of data
streaming applications, we are interested in deterministic samples
that can be constructed using space that is polylogarithmic in the
data stream's length.

In addition, because much of the streaming data 
is coming from sensors and scientific instruments, we are interested
in this paper in studying streaming algorithms for geometric data.
Such data could include multi-dimensional points in the color space
of astrophysical data or two-dimensional lines defined 
by a point-line duality of a stream of points in the plane.
Of particular interest, then, is data streaming algorithms for
constructing \e-nets and \e-approximations, which are general
structures developed in the computational geometry literature for
deterministically sampling geometric data. 
Indeed, \e-nets and \e-approximations are developed in a very general
context of bounded-dimensional range spaces, where we are given a
ground set and a polynomial-sized family of ranges on that set (which
constitute the queries or sampling statistics we are interested in).
Hence, results for constructing such deterministic samples should
have a considerable number of applications.

\subsection{Related work on Streaming Algorithms} 
Data streaming problems have engendered a large amount of interest
among the algorithms community over the last few years. For a
comprehensive survey of the work done so far and some interesting
directions for the future, the reader is referred to Muthukrishnan's
work~\cite{muthu-survey-03}. An earlier survey by Babcock
et. al.~\cite{babcock-models_streams-02} explores the issues arising
in building data stream systems. 

We are not familiar with any previous work for constructing \e-nets
or \e-approximations in streaming models, although these structures
have been extensively studied in full-memory contexts
(e.g., see the chapter by Matou\v{s}ek \cite{m-dcg-00}).
The closest previous work is done 
in the {\em iceberg query}~\cite{fang-iceberg-98} framework of
Manku and Motwani~\cite{manku-frequency-02}, who provide $1+\epsilon$
approximations for the frequency counts of items in a data stream
that occur more than $\epsilon N$ times (which are the so-called ``icebergs'').
Similarly, algorithms for
computing the quantiles of a data stream have been given by Greenwald
and Khanna~\cite{greenwald-quantiles-01} guaranteeing a precision of
$\epsilon N$, which is similar to the guarantees that are
provided by \e-approximations, while using
$O(\frac{1}{\epsilon} \log \epsilon N)$ space. This limitation of an
additive $\epsilon N$ error in every quantile is overcome by Gupta and
Zane~\cite{gupta-inversions-03}. The latter's method provides relative
error for all quantiles but uses $O(\log^2 N / \epsilon^3)$ space and
requires knowledge of an upper bound on the stream size.

The first geometric problem to be studied in the streaming model was
that of finding the diameter of a set of points. Feigenbaum, Kannan
and Zhang~\cite{feigenbaum-diameter-02} gave an $O(1/\epsilon)$ space
algorithm for computing the diameter of points in two dimensions in
the streaming model and a 
$O(\frac{1}{\epsilon^{3/2}} \cdot \log^3 N 
  (\log R + \log\log N + \log(\frac{1}{\epsilon})))$ space algorithm for
computing it in the sliding window model where $R$ is the maximum,
over all windows, of the ratio of the diameter to the distance between
the closest two points in the window. Indyk~\cite{indyk-diameter-03}
gave a streaming algorithm which maintains a $c$-approximate diameter
of points in $d$ dimensions using $O(dn^{1/(c^2-1)})$ space taking
$O(dn^{1/(c^2-1)})$ time per new point, for $c > \sqrt{2}$.

Cormode and Muthukrishnan generalized the exponential histograms used
on single dimensional data sets in earlier works on streaming
algorithms~\cite{datar-statistics-02,korn-histograms-02} and defined
{\em radial histograms}~\cite{cormode-radial-03}, which allowed them to
give a $O(1+\epsilon)$ approximation to the diameter using
$O(1/\epsilon)$ space.  They were also able to use these structures to
approximate convex hulls in the sense that no point in the input
stream is more than $\epsilon D$ outside the approximate hull, where
$D$ is the diamter of the point set. Constructing an approximate hull
takes them $O(q/\epsilon)$ space. Hershberger and
Suri~\cite{hershberger-diameter-03} improve this to
give a sampling-based algorithm
for approximating the convex hull of a streaming point set, showing 
how to maintain an adaptive sample of at most $2r$ points such
that the distance between the hull of their sample and the true convex
hull is $O(D/r^2)$, where $D$ is the current diameter of the sample.
Some of the other geometric problems that have been studied in a
streaming model include minimum spanning tree and minimum weight
matching~\cite{indyk-mst-03} and certain facility location and nearest
neighbour kind of queries~\cite{cormode-radial-03}.

\subsection{Our Results}
In this paper, we present memory-efficient deterministic algorithms
for constructing \e-nets and \e-approximations of 
streams of geometric data.
Our algorithms use a polylogarithmic
amount of memory, provided $\epsilon$ is at least inverse-polylogarithmic.
As mentioned above, \e-nets and \e-approximations are of interest in
their own right and have many applications in computational geometry.
We show how our deterministic samples can be used to answer
online iceberg geometric queries on data streams, such as in
multi-dimensional iceberg range searching.
Because the information typically of interest from data streams is
statistical, we focus in this paper primarily on the use of \e-nets and
\e-approximations to compute approximations to several
robust statistics of geometric data streams, including Tukey depth, simplicial
depth, regression depth, the Thiel-Sen estimator, and the least
median of squares.
Thus, we additionally give polylogarithmic-space data streaming
algorithms for computing approximations to these statistics.
We also include a lower bound for non-iceberg range queries in data
streams.

%-----------------------------------------------------------
\section{Preliminaries on \e-Nets and \e-Approximations}

In this section we recap certain aspects of \e-Nets and
\e-approximations~\cite{vc-ucrfe-71,m-dcg-00}, which are part of a
general framework for modelling a number of interesting problems in
computational geometry and derandomizing divide-and-conquer type
algorithms.

A {\em range space} is a set system, i.e., a pair $\Sigma = (X,\R)$,
where $X$ is a set and \R\ is a set of subsets of $X$.  We call the
elements of \R\ the {\em ranges} of $\Sigma$, as \R\ is typically
defined in terms of some well structured geometry.  If $Y$ is a subset
of $X$, we denote by $\R|_Y$ the set system {\em induced by \R\ on
$Y$}, i.e., $\{R \cap Y | R \in \R \}$\footnote{Note that although
many sets of \R\ may intersect $Y$ in the same subset, this
intersection appears only once in $\R|_Y$.}.

We say a subset $Y \subseteq X$ is {\em shattered} if every possible
subset of $Y$ is induced by \R, i.e., if $\R|_Y = 2^Y$.  The {\em
VC-dimension} of $\Sigma$ is the maximum size of a shattered subset of
$X$.  If there are shattered subsets of any size, then the
VC-dimension is infinite.  A related and simpler notion is the 
{\em scaffold dimension}~\cite{gr-bidgp-97} of $\Sigma$.  
It is based on the notion of the
{\em shatter function} $\pi_{\R}(m)$, which we define as the maximum
possible number of sets in a subsystem of $\Sigma$ induced by an
$m$-sized subset of $X$.  In other words, it is the $\sup\{|R|_Y | : Y
\subseteq X, |Y| = m\}$.  We now define the scaffold dimension of
$(X,\R)$ as the infimum of all numbers $d$ such that $\pi_{\R}(m)$ is
$O(m^d)$.  It turns out that the shatter function of a set system of
VC-dimension $d'$ is bounded by $\binom{m}{0} + \binom{m}{1} + \cdots
+ \binom{m}{d'} = \Theta(m^{d'})$ \cite{s-dfs-72, vc-ucrfe-71}.  Thus
the scaffold dimension is always at most the VC-dimension.
Conversely, if the scaffold dimension is bounded by a constant, the
VC-dimension too is bounded by a constant.  There are, however, many
natural geometric set systems of scaffold dimension strictly smaller
than the VC-dimension; for instance, the scaffold dimension of a set
system defined by halfplanes in the plane is 2, while the VC-dimension
is 3.  In the rest of the paper, we will always refer to the scaffold
dimension of a set system.  In addition, we consider only those set
systems whose scaffold dimensions are bounded by a constant.

We are now ready to define \e-nets and \e-approximations.  A subset $S
\subseteq X$ is an {\em \e-net for $(X,\R)$} provided that $S \cap R
\not = \emptyset$ for every $R \in \R$ with $|R|/|X| < \e$.  A subset
$A \subseteq X$ is an {\em \e-approximation for $(X,\R)$} provided
that
\begin{equation} \label{eq:e-approx}
\left | \frac{|A \cap R|}{|A|} - \frac{|X \cap R|}{|X|} \right | \leq \e
\end{equation}
for every set $R \in \R$.  Note that every \e-approximation is
automatically an \e-net, but the converse need not be true.  A
remarkable property about set systems of scaffold dimension $d$ is
that, for any $\e \in [0,1)$, they admit an \e-approximation whose
size depends only on $d$ and \e, {\em not} on the size of $X$.  The
first basic result in this vein is the following lemma.
\begin{lemma}
   For any set system $(X,\R)$, with a finite $X$, and a scaffold
   dimension at most $d$, where $d \geq 1$, there exists, for any $\e
   \in [0,1]$, an \e-net of size at most $C_{1}\e^{-1} \lg(\e^{-1})$, and
   an \e-approximation of size at most $C_{2}\e^{-2} \lg(\e^{-1})$.
   Here $C_1, C_2$ depend on only $d$.
%   For any $d \geq 1$ there exists a $C(d)$ such that for any $\e \in
%   [0,1]$ and for any set system $(X,\R)$ with $X$ finite and of
%   dimension at most $d$ there exists an \e-net of size at most
%   $C(d)\e^{-1} \lg(\e^{-1})$.  In fact, a random sample $S \subseteq
%   X$ of this size is an \e-net with a positive probability.
\end{lemma}
Note that, in general, the $\lg(\e^{-1})$ factor cannot be removed
from the bound.

Matou\v{s}ek \cite{m-aogdc-95} gave a deterministic algorithm for
efficiently computing small sized \e-approximations (and thereby,
\e-nets) for set systems with constant-bounded scaffold dimensions.
Such an algorithm needs that the set system to be given in a form more
``compact'' than simply the listing of the elements in each set. For
this we assume the existence of a {\em
subsystem oracle}, i.e. an algorithm (depending on
the specific geometric application) that, given any subset $Y
\subseteq X$, lists all sets of $\R|_Y$.  We say that the subsystem
oracle is {\em of dimension at most $d$} if it lists all sets in time
$O(|Y|^{d+1})$.  This corresponds to the scaffold dimension; the
maximum number of sets in $\R|_Y$ is $\pi_{\R}(|Y|)$, and the ``$+1$''
in the exponent accounts for the fact that each output set is given by
a list of size up to $|Y|$.  Matou\v{s}ek's result is summarized 
by the following lemma.
\begin{lemma}\label{lm:e-approx}
  Let $(X,\R)$ be a set system with a subsystem oracle of dimension
  $d$, where $d$ is a constant.  Given any $\e \in [0,1)$, we can
  compute an \e-approximation of size $O(\e^{-2}\lg(\e^{-1}))$ and an
  \e-net of size $O(\e^{-1}\lg(\e^{-1}))$ in time
  $O(|X|\e^{-2d}\lg^{d}(e^{-1}))$.
\end{lemma}
We shall use the algorithm above as a sub-routine for our streaming
algorithm for \e-approximations (see Section \ref{sec:algorithm}).  It
is based on two observations that we state below. They correspond to
two basic operations of our algorithm, the {\em merge step} and the
{\em reduce step}.  Many algorithms for computing \e-approximations
(certainly the one Matou\v{s}ek gave, and the one we shall give) start
by partitioning $X$ into small pieces, and then essentially alternate
between the two steps until they get the desired approximation.

\begin{observation}[Merge Step] \label{ob:merge}
  Let $X_1, \ldots, X_m \subseteq X$ be disjoint subsets of equal
  cardinality and let $A_i$ be an \e-approximation of cardinality $b$
  for $(X_i, \R|_{X_i}), i = 1, \ldots, m$.  Then $A_1 \cup \ldots
  \cup A_m$ is an \e-approximation for the subsystem induced by \R\ on
  $X_i \cup \ldots \cup X_m$.
\end{observation}

\begin{observation}[Reduce Step] \label{ob:reduce}
  Let $A$ be an \e-approximation for $(X,\R)$ and let $A'$ be a
  $\delta$-approximation for $(A,\R|_A)$.  Then $A'$ is an $(\e +
  \delta)$-approximation for $(X,\R)$.
\end{observation}
%% We would have ended this introduction to \e-approximations here, but
%% for that we need stronger versions of both Lemma \ref{lm:e-approx}
%% and Observation \ref{ob:merge} in order to describe the streaming
%% algorithm.  These stronger versions, both, are generalization that
%% involve weights, though the manner in which weights are used is
%% slightly different.

Before we end this preliminary section, we state the following
extension to Lemma \ref{lm:e-approx}.  This, too, was given by
Matou\v{s}ek \cite{m-aogdc-95}.

\begin{lemma} \label{lm:weighted_e-approx}
  Let $X$ be a finite set equipped by a probabilistic measure $\mu$
  (given by a table) and let $\Sigma = (X,\R)$ be a range space
  satisfying the assumptions of Lemma \ref{lm:e-approx}.  Then an
  \e-approximation for $\Sigma$ with respect to the measure $\mu$ can
  be computed with the same asymptotic efficiency in the running time
  and size of the \e-approximation in the case of uniform measure in
  Lemma \ref{lm:e-approx}.
\end{lemma}
When $X$ is associated with a probabilistic measure $\mu$, an
\e-approximation of $(X,\R)$ is a multi-set $A$ such that
\[
\left | \frac{|A \cap R|}{|A|} - \frac{\mu(X \cap R)}{\mu(X)} \right | \leq \e
\]
for every $R \in \R$.

\section{Some Additional Extensions for Weighted Sets}
While the extensions described above are useful in our context, we
nevertheless need some further generalizations, which will be useful
in the data streaming model.  In particular, we need to generalize
Observation~\ref{ob:merge} to a weighted case.  This allows us to
merge \e-approximations of different sizes and for sets of different
cardinalities.  To the best of our knowledge, this is the first time
such an observation is being made.  Note that in the un-weighted case,
for an \e-approximation $A$ for $(X,\R)$, each element in $A$
``represents'' $|X|/|A|$ elements in $X$.  This is easy to see if we
write Requirement \ref{eq:e-approx} in the following form
\[
\left | |A \cap R|\frac{|X|}{|A|} - |X \cap R| \right | \leq \e|X|
\]
Now, instead of having an element p in the \e-approximation $A$ represent
the same number of elements in $X$, we can assign it a weight $\gamma(p)$
equal to the number of elements in $X$ that it represents.  In this
generalized scenario, a subset $A \subseteq X$, is a {\em weighted
\e-approximation for $(X,\R)$} provided that
\[
\left | \sum_{p \in A \cap R} \gamma(p) - |X \cap R| \right | \leq \e|X|.
\]
We are now ready to state our observation related to weighted merging.
\begin{observation}[Weighted Merge Step] \label{ob:weighted_merge}
  Let $X_1, \ldots, X_m \subseteq X$ be disjoint subsets (of
  cardinalities not necessarily the same) and let $A_i$ be a weighted
  \e-approximation of $(X_i, \R|_{X_i}), i = 1, \ldots, m$.  Then $A_1
  \cup \ldots \cup A_m$ is a weighted \e-approximation for the
  subsystem induced by \R\ on $X_i \cup \ldots \cup X_m$, where the
  weights on the points remain as they were.
\end{observation}

%This ends our introduction to \e-nets and \e-approximations.  We are
%now ready to describe our streaming algorithm in the next section.

%=====================================================================
%  Algorithm
%=====================================================================

\section{Computing \e-Approximations in Geometric Streams} 
\label{sec:algorithm}

% \comment{ David's Notes: Anyway, the details of the
% $\epsilon$-approximation algorithm: This is in the insertion-only model,
% so let $p_i$ denote the $i$th inserted object.  Define a canonical set to
% be a subset of already-inserted objects of the form $\{ p_i | k 2^j <= i
% < (k+1) 2^j \}$, where $j$ and $k$ are nonnegative integers, and identify the
% set by the pair $(j,k)$.  Each canonical set $(j,k)$, $k > 0$, has two
% children $(2j,k-1)$ and $(2j+1,k-1)$.  Define a maximal canonical set to
% be a set whose parent is not yet canonical (because not all its
% members have been inserted yet). Let $w_i$ be a series where
% $W=\sum_{i=0}^{\infty}w_i=O(1)$.  Whenever a set $(j,k)$ becomes canonical,
% we compute an $(\epsilon*w_i/2W$)-approximation to the union of the
% approximations previously computed for its two children.  By the
% compositional properties of \e-approximations, this is a
% $(\sum_{i=k}^{\infty}\epsilon*w_i/2W)$-approximation (and therefore an
% $\epsilon/2$-approximation) to the whole canonical set.  We store in our
% data structure these approximations for the current maximal canonical
% sets, and compute after each step an $(\epsilon/2)$-approximation to the
% union of these stored approximations.  Letting e.g. $w_i = i^{1+1/2c}$
% we get total space $O(\epsilon^{-2-1/c})$ for any constant $c$.}

Let $x_1,\ldots,x_n,\ldots$ be a stream of geometric objects in the
time series model. Let $X$ be the set of all the objects in the stream
that have arrived till now.  Let \R\ be a set of ranges defined on
$X$, and $\Sigma = (X,\R)$ be the current range space.  In addition,
let $d$, where $d$ is a constant, be the scaffold dimension of
$\Sigma$.

We present an algorithm which computes an \e-approximation to
$(X,\R)$, given any $\e \in [0,1)$. Our algorithm maintains a
polylogarithmic-sized data structure from which it computes this
\e-approximation. Additionally, it takes polylogarithmic time to
update this structure on the arrival of a new item in the stream.

The \e-approximation our algorithm produces is asymptotically equal in
size to that produced in the static model (see Lemma
\ref{lm:e-approx}).  Interestingly, our algorithm does not need to
know the value of $n$ in advance.  Our algorithm simulates the
divide-and-conquer approach of the static algorithm in a bottom up
fashion. We now outline how this is done.

We begin by imposing a hierarchy of groupings onto the stream: define
{\em canonical sets} $S_{j,k}$ as $\{x_i | j 2^k \leq i < (j+1) 2^k
\}$ for $j, k \geq 0$.  Canonical sets are inter-related through a
natural tree hierarchy.  The {\em children} of set $S_{j,k}$, $k \geq
1$, are the canonical sets $S_{2j,k-1}$ and $S_{2j+1,k-1}$.  We say
that a canonical set $S_{j,k}$ becomes {\em available} when the last
element in it, i.e., $x_{(j+1)2^k -1}$, arrives.  A {\em maximal
canonical set} is one that is available but whose parent is not yet
available.  Observe that when $x_n$ arrives, there are at most $\lg n$
maximal canonical sets.  Also, the union of all the maximal canonical
sets is the set $X$ of all elements that have arrived till now.

We use the following building blocks.
\begin{itemize}
\item \eapprox: An algorithm for deterministically computing
  \e-approximation with small size (see Lemma~\ref{lm:e-approx}.)

\item \weapprox: An algorithm for
  deterministically computing \e-approximations of {\em weighted}
  items (see Lemma~\ref{lm:weighted_e-approx}.)
\end{itemize}

Note that we cannot afford to use \eapprox\ on an input that is larger
than logarithmic, as otherwise we will not remain within our space and
time bounds.  

Our algorithm, we call it \esapprox, follows the basic merge and
reduce technique~\cite{m-dcg-00} for constructing
\e-approximations. To follow this technique we need to use a sequence
$w_1,\ldots,w_u,\ldots$ with the property that $W \triangleq
\sum_{u=1}^{\infty} w_u = O(1)$.  Here we shall use $w_i = i^{-1-c}$,
for some $c > 0$.

\begin{figure}[h]
\centering
\includegraphics[scale=0.6]{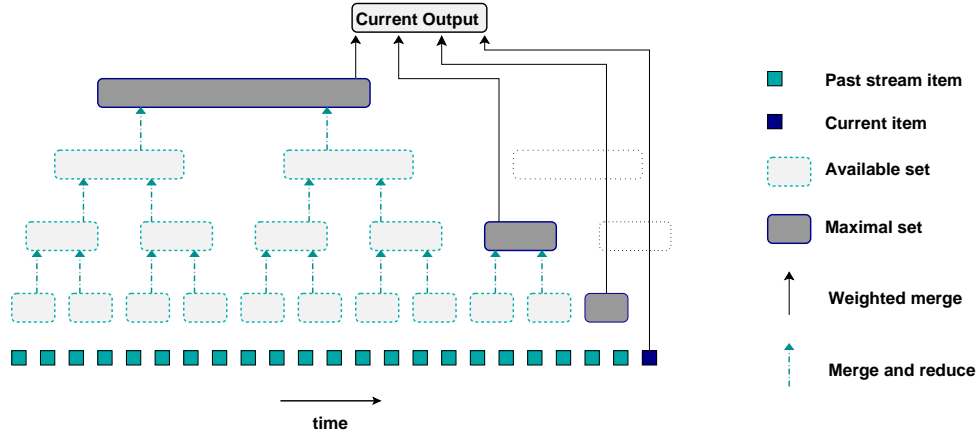}
\caption{Schematic: Computing an $\epsilon$-approximation of a data stream}
\label{fig:algorithm}
\end{figure}

At a high level the algorithm is as follows (see
Figure~\ref{fig:algorithm}): At every stage, the algorithm stores a
$\delta$-approximation for all available maximal canonical sets, where
$\delta$ varies with the set, but is always at most $\e/2$.  Let
$A_{j,k}$ be such an approximation for $S_{j,k}$. This
$\delta$-approximation is constructed through merging the
approximations $A_{2j,k-1}$ and $A_{2j+1,k-1}$ which were earlier
computed for $S_{j,k}$'s two children.

The \e-approximation of the set $X$ at any point, the {\em stream
output}, is determined by weighted merging.  Each element $p \in
A_{j,k}$ is assigned a weight $\gamma(p) = |S_{j,k}|/|A_{j,k}|$ for
this purpose.  As it happens, once a weight is assigned to an object,
we don't ever need to change it.

\begin{figure}[h]
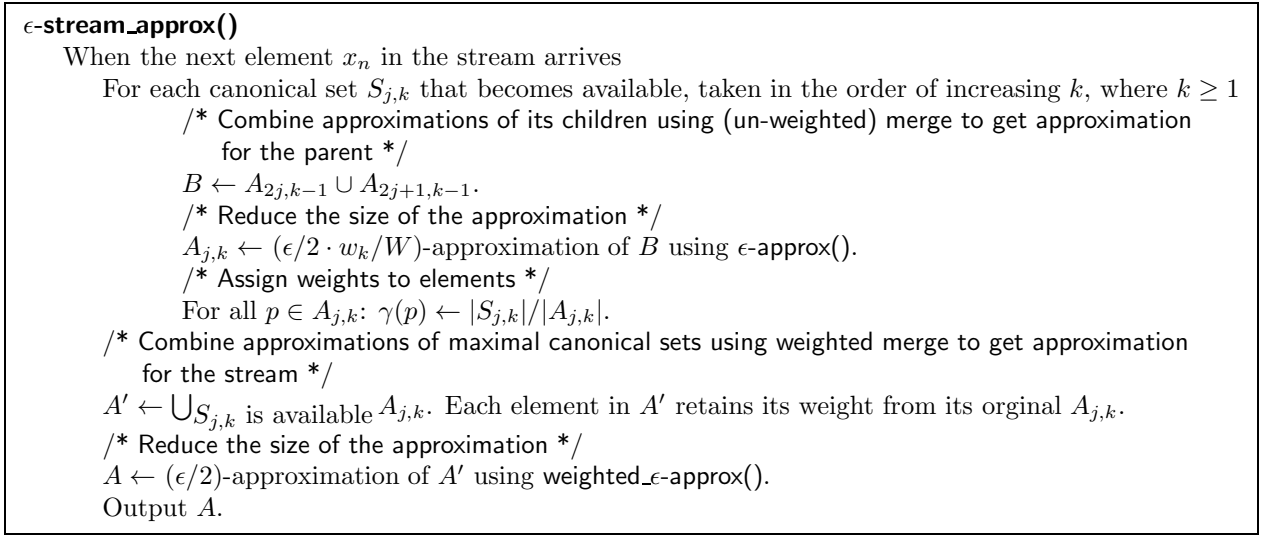

\begin{center}
\fbox{
\begin{minipage}{\textwidth}
\small
\begin{tabbing}
012\=012\=012\=012\=012\=012\=012\=012\=012\=012\=012\=012\= \kill
{\bfseries \esapprox}\\ \>When the next element $x_n$ in the stream
arrives\\

\> \>For each canonical set $S_{j,k}$ that becomes available, taken in the
order of increasing $k$, where $k \geq 1$\\

\>  \> \> \>\algocomment{/* Combine approximations of its children using
  (un-weighted) merge to get approximation}\\
\>  \> \> \> \> \algocomment{for the parent */}\\
\>  \> \> \>$B \leftarrow A_{2j,k-1} \cup A_{2j+1,k-1}.$ \\

\>  \>  \>  \>\algocomment{/* Reduce the size of the approximation */}\\
\>  \>  \>  \>$A_{j,k} \leftarrow$ $(\e/2 \cdot w_k/W)$-approximation of
$B$ using \eapprox. \\

\> \> \> \>\algocomment{/* Assign weights to elements */}\\ \> \> \>
\>For all $p \in A_{j,k}$: $\gamma(p) \leftarrow |S_{j,k}|/|A_{j,k}|.$ \\

\>  \>\algocomment{/* Combine approximations of maximal canonical sets
  using weighted merge to get approximation}\\
\>  \>  \>\algocomment{for the stream */}\\
\> \>$A' \leftarrow \bigcup_{\mbox{$S_{j,k}$ is available}} A_{j,k}$.
  Each element in $A'$ retains its weight from its orginal
  $A_{j,k}$. \\

\>  \>\algocomment{/* Reduce the size of the approximation */}\\  
\>  \>$A \leftarrow$ $(\e/2)$-approximation of $A'$ using \weapprox.\\
\>  \>Output $A$.
\end{tabbing}
\end{minipage}
}
\end{center}
\caption{Algorithm for computing an \e-approximation of a geometric
  stream.}
\label{fig:algorithm_description}
\end{figure}

We are now ready to formally specify \esapprox.  Figure
\ref{fig:algorithm_description} contains the specification. Assume
that $A_{j,0}$ is the element itself in the singleton set $S_{j,0}$.

\paragraph{Correctness, Space, and Time.} Observations
\ref{ob:merge} and \ref{ob:reduce} imply that $A_{i,j}$ is a
$\delta$-approximation for $S_{j,k}$, where
\[
\delta \leq \sum_{u=1}^{k} \frac{\e}{2} \cdot \frac{w_u}{W} <
\frac{\e}{2}.
\]
Together with Observation \ref{ob:weighted_merge}, this implies that
$A'$ is a weighted $(\e/2)$-approximation for the set $X$ of elements
in the stream.  Now bring in the properties of \weapprox\ and another
application of Observation \ref{ob:reduce} to see that $A$ is indeed
an \e-approximation of $(X,\R)$.

The data structure needs to store just the `$A_{j,k}$'s; all other
sets are intermediate results that can be discarded.  Lemma
\ref{lm:e-approx} implies that the size of $A_{j,k}$, is $O((\e \cdot
w_k)^{-2} \lg((\e \cdot w_k)^{-1}))$; remember that the size is
determined by just the last reduction step.  Denote the size of
largest such set, i.e., $A_{j,\lg n}$, by $s = s(n,\e^{-1})$, which is
$O(\lg^{2+2c}n \cdot \e^{-2} \cdot (\lg\lg n - \lg\e))$.  Note that
$s$ is also an upper bound for the size of the input to \eapprox.

Consider the space and time requirements.  These are dominated by the
requirements for \weapprox.  The input to \weapprox is $O(\lg n \cdot
s)$.  Lemma \ref{lm:weighted_e-approx} implies that the space and time
requiments for \esapprox\ are $O(\lg^{3+2c}n \cdot
e^{-2d-2}\lg(e^{-1}) \cdot (\lg\lg n - \lg \e))$.

%% Consider the space needed.  Using again the result
%% in Lemma \ref{lm:e-approx} \eapprox\ takes
%% $O(\lg^{2+2c} n \cdot \e^{-2d-2} \lg(\e^{-1}) \cdot (\lg\lg n -
%% \lg\e))$ space (and time). At any stage there are at most $\lg n$
%% approximations.  Thus space is $O((\lg^{2+2c} n \cdot \e^{-2d-2}
%% \lg(\e^{-1}) + \lg^{3+2c} n \e^{-2}) \cdot (\lg\lg n - \lg\e))$.
%% After a stream element arrives, \eapprox\ may run at most $\lg n$
%% times; the time taken is $O(\lg^{3+2c} n \cdot \e^{-2d-2} \lg(\e^{-1})
%% \cdot (\lg\lg n - \lg\e))$.

%=====================================================================
%  Applications
%=====================================================================
\section{Applications} 
\label{sec:applications}

\e-Nets and \e-approximations have a number of applications in
computational geometry, and even other areas like learning theory ---
see, e.g., \cite{m-encg-93}.  Many of the problems in these have
streaming versions.  One basic application is {\em range counting}.
In this, we are given a set $S$ of $n$ points in $\mathbb{R}^d$, and a
family \R\ (the ranges) of subsets of $\mathbb{R}^d$.  Each query
consists of a range $R \in \R$ and asks for the number of points in
it.  Typical range families are axes-orthogonal ranges, spherical
ranges (proximity queries), and simplical ranges.  The corresponding
range spaces for these all have a bounded scaffold dimension.  In the
streaming version, the point set $S$ comes as a continuous stream,
interspersed with queries.  It is easy to see how our algorithm would
work here: use \esapprox\ to maintain an \e-approximation $A$ of the
current $(S,\R)$.  When queried with range $R \in \R$, output $|A \cap
R| \cdot n/|A|$.  This is within an additive $\e n$ of the true value;
this is akin to the iceberg queries mentioned earlier.

The above technique has implications in a lot of specific
applications.  To get a flavor of this, we delve deeper into the
specific area of robust statistic in the next few paragraphs.

\subsection{Robust Statistics}

{\em Robust statistics} concerns the study of statistical estimators that can tolerate high numbers of
{\em outliers}, while maintaining an accuracy of estimation that depends only on the remaining uncorrupted data points.  In contrast, ordinary least squares estimators, while trivial to compute even in the streaming model, can be forced to produce estimates that are arbitrarily far from the correct model even in the presence of a single outlier.
The number of outliers that an estimator can tolerate while preserving its accuracy is called its {\em breakdown point}; in general, methods with high breakdown points are preferred but other criteria are also important including statistical efficiency (number of samples needed to achieve a given accuracy) and computational efficiency (amount of time it takes to compute a given estimate from a set of samples).  Many robust statistical methods also have the advantage of being {\em non-parametric}, not requiring the statistician to produce a prior probability distribution or other arbitrary parameters before producing a fit.  The paradigmatic example of a robust statistic is the median of one-dimensional data, which, unlike the mean, is robust with a breakdown point of $\frac12$.
Much research on streaming algorithms has gone into methods for maintaining approximate medians or more general quantiles~\cite{greenwald-quantiles-01}, and we would like to find similar methods for higher dimensional statistics.

Two of the critical problems studied in robust statistics are {\em location} (finding a central point in a cloud of data points) and {\em regression} (fitting the data to a model in which a dependent variable or variables is a linear function of the independent variables).
Many methods in this area are based on various concepts of {\em depth}, which measures the quality of fit of an estimate.  It is natural to seek the estimate maximizing the depth,
but it is also of importance to be able to compute depths of non-optimal estimates,
in order to form {\em depth contours} that produce a center-outward ordering of the data.

For many of these robust statistical methods, a computationally efficient streaming approximation to the depth measure can be obtained from an $\epsilon$-approximation of the sample data.
The deepest fit can be approximated by a deepest fit to the $\epsilon$-approximation,
and this approximate fit often has similar breakdown point properties to the non-approximate fit on which it is based.  We describe below several of the methods to which this technique applies:

\paragraph{Tukey Depth.}
This quantity~\cite{donoho-bpmle-82} measures the quality of fit of a
center, as the minimum proportion of sample points among all halfspaces
that contain the center.  The Tukey depth of a point can be computed in
time $O(n^d\log n)$, where $n$ denotes the number of sample
points~\cite{rousseeuw-compdepth-98}.  The {\em Tukey median} is the
point of maximum depth.  It is known that any Tukey median has depth at
least $1/(d+1)$, and the breakdown point of the Tukey median as an
estimate of location is also $1/(d+1)$.  There are known static
algorithms for finding Tukey medians, or other points of high depth, in
two or three
dimensions~\cite{jadhav-centerpoint-93,langerman-optarr-03,matousek-center-91},
but in higher dimensions only inefficient linear-programming based
exact solutions are known and it is necessary to resort to more
efficient approximation algorithms~\cite{clarkson-approxcenter-96}.

Since the Tukey depth is based on counting points in halfspaces, it can
be approximated using $\epsilon$-approx\-i\-ma\-tions for halfspace
ranges~\cite{clarkson-approxcenter-96}:  the depth of a point within an
$\epsilon$-approximation of a sample is within an additive error of
$\epsilon$ of its depth in the original sample data.  In particular,
the Tukey median of an $\epsilon$-approximation has depth within
$\epsilon$ of that of the true Tukey median.  The breakdown point of
this approximate Tukey median is $1/(d+1)-\epsilon$.  Thus, by using
our streaming $\epsilon$-approximation algorithm, we can efficiently
maintain not only an approximate Tukey median of the data set, but also
a space-efficient data structure from which we can compute accurate
approximations of the Tukey depth of any point.

\paragraph{Simplicial Depth.} 
This is another measure of quality of fit for location, introduced by
Liu~\cite{liu-simplexdepth-90}.  The simplicial depth of a fit point is
defined to be the proportion of simplices, among all the $n\choose d+1$
simplices formed by convex hulls of $(d+1)$-tuples of sample points,
that contain the fit point.  Equivalently, it is the probability that a
randomly chosen $(d+1)$-tuple contains the fit point in its convex
hull.  As we now argue, for points in the plane, the simplicial depth
in a sample set is accurately approximated by the simplicial depth of
an $\epsilon$-approximation for wedge ranges (that is, ranges formed by
intersecting two halfplanes).  Therefore, as for Tukey depth, we can
answer approximate depth queries and maintain an approximate deepest
point in a space-efficient manner for streaming data.

Let $\delta$ be a value to be determined later and imagine the
following process for measuring approximately the simplicial depth of a
fit point: first, let $L$ be a set of $1/\delta$ lines through the fit
point, partitioning the plane into $2/\delta$ wedges having the fit
point as a common apex, with at most a $\delta$ fraction of the sample
points in any wedge.  Let $e_1$ be the proportion of triangles,
determined by three input points, that are not all on one side of one
of a line in $L$.  Then $e_1$ is an overestimate of the simplicial
depth, but the amount by which it overestimates the depth is
$O(\delta)$: the only triangles incorrectly included in the estimate
are ones that have two points in opposite wedges, there are $O(\delta^2
n^3)$ such triangles per pair of opposite wedges, and $O(1/\delta)$
such pairs.  Next, let $e_2$ be the proportion of triangles, determined
by three points in an $\epsilon$-approximation of the sample, that are
not all on one side of a line in $L$.  For the same reasons as before,
$e_2$ is within $O(\delta)$ of the simplicial depth for the
$\epsilon$-approximation.  Further, $e_1$ and $e_2$ are within
$O(\epsilon/\delta)$ of each other:  $$e_1=1 - \sum_i {{w_i\choose 3} +
{w_i\choose 2}(h_i-w_i) + w_i{h_i-w_i\choose 2}\over {n\choose 3}},$$
where $w_i$ is the number of sample points in the $i$th wedge and $h_i$
is the number of sample points in the halfplane containing the $i$th
wedge on its counterclockwise boundary.  Each term in the sum is
approximated within $O(\epsilon)$ by the corresponding term where $w_i$
and $h_i$ are replaced by numbers of points in the
$\epsilon$-approximation, and there are $O(1/\delta)$ terms, so the
total difference between $e_1$ and $e_2$ is $O(\epsilon/\delta)$.
Putting together the errors in going from the original simplicial depth
to $e_1$ to $e_2$ to the simplicial depth of the approximation, and
setting $\delta=\sqrt\epsilon$, we see that the
$\epsilon$-approximation approximates the simplicial depth to within
$O(\sqrt\epsilon)$.

As far as we are aware, this deterministic $\epsilon$-approximation
based method for approximating simplicial depth is novel even for
static, non-streaming data, although it is trivial to approximate
simplicial depth randomly in the static case by sampling triangles.  It
seems likely that similar deterministic and streaming approximation
guarantees, with worse dependence on $\epsilon$, can be shown to hold
also in higher dimensions.

\paragraph{Regression Depth.} 
This statistic
was introduced by Rousseeuw and Hubert~\cite{rousseeuw-regdepth-99} as
a measure of the quality of fit of a regression hyperplane.  It is
defined as being the minimum proportion of sample points that can be
removed to turn the fit plane into a {\em nonfit}, that is, a
hyperplane combinatorially equivalent to a vertical hyperplane.  Amenta
et al.~\cite{amenta-regdepth-00} showed that, like Tukey depth, for
regression depth a fit always exists with depth at least $1/(d+1)$, and
the breakdown point of the maximum-depth fit is $1/(d+1)$.  Their proof
technique shows that the regression depth of a query hyperplane can be
measured by performing a certain projective transformation of the space
containing the sample points, and measuring the Tukey depth of a
certain point in the transformed space.   Due to the transformation, a
halfspace in the transformed space may correspond to a {\em double
wedge} (symmetric difference of two halfspaces) in the original space.
Therefore, the same $\epsilon$-approximation technique used for Tukey
depth, but with double wedge ranges, also applies to regression depth,
and lets us compute depths and maintain an approximate deepest fit with
high breakdown point for streaming data.  Bern and
Eppstein~\cite{bern-multivariate-02} generalized regression depth to
the context of multivariate regression, in which the sample data have
more than one dependent variable; in their definition, the depth of a
fit is the minimum proportion of sample data contained in any double
wedge, one boundary of which contains the fit and the other of which is
parallel to the dependent coordinate axes; this is again well
approximated by $\epsilon$-approximations for double wedge ranges.

\paragraph{The Thiel-Sen Estimator.}
This estimator~\cite{sen-regress-68,thiel-regress-50} is a method for
two-dimensional linear regression, in which one first finds the median
among all $n\choose 2$ slopes determined by the lines through pairs of
sample points, and then selects a regression line having that median
slope and bisecting the sample set.  It has a breakdown point of
$1-\sqrt{1/2}\approx 0.293$.  This has long been a testbed for
geometric optimization algorithms, and several $O(n\log n)$ time static
algorithms for it are known, among them one based on using
$\epsilon$-cuttings in a prune-and-search
technique~\cite{bronnimann-slopesel-98}.  However these algorithms seem
to require repeatedly scanning the data in a way that is unavailable to
a streaming algorithm.  Instead, we apply an approximation technique
very similar to that for simplicial depth, above.

To begin with, suppose that we are given a query slope $s$, and must
determine the approximate position of $s$ within the sorted sequence of
slopes, normalized by dividing the position by $n\choose 2$.  This can
be solved exactly by a reduction to computing the number of inversions
in a permutation, but we are interested in approximations that can be
computed by a streaming algorithm that does not know $s$ in advance.
To do this, let $\delta$ be a parameter to be determined later, and
imagine subdividing the sample points into a grid by $O(1/\delta)$
lines that are vertical and parallel to $s$, in such a way that at most
a $\delta$ proportion of the points lie in the slab between any two
adjacent parallel grid lines.  Let $e_1$ be an estimate of the position
of $s$, formed by summing up the normalized number of pairs of points
that form a line with lower slope than $s$ and that are in a pair of
grid cells that are separated both by a vertical line of the grid and
by a line parallel to $s$ from the grid.  Then $e_1$ is within
$O(\delta)$ of the true position of $s$ since the only lines through a
given point that are omitted from the count are the ones where the
other point determining the line is in one of the two slabs containing
$s$, and $e_1$ can be expressed as a sum with $O(\delta^{-2})$ terms,
each term being a product of the number of points in two
parallelograms.  Let $e_2$ be a similar normalized sum, with the number
of sample points in each parallelogram replaced by the number of points
of an $\epsilon$-approximation for parallelogram ranges, and let $e_3$
be the normalized position of $s$ within the set of lines determined by
pairs of points from the $\epsilon$-approximation.  Then $e_1$ differs
from $e_2$ by $O(\epsilon\delta^{-2})$ and $e_2$ differs from $e_3$ by
$O(\delta+\epsilon\delta^{-1})$.  Therefore, the overall error caused
by using $e_3$ as our approximation to the position of $s$ is $O(\delta
+ \epsilon\delta^{-2})$.  Setting $\delta=\epsilon^{1/3}$ makes this
total error equal $O(\epsilon^{1/3})$.

To compute an approximate Thiel-Sen estimator, we use the same
$\epsilon$-approximation for parallelograms, compute the median slope
among pairs of points from the approximation, and then find a line with
that median slope bisecting the approximation.  The resulting line has
slope with a normalized position within $O(\epsilon^{1/3})$ of the
median slope, partitions the sample points within $\epsilon$ of exact
bisection, and has a breakdown point of
$1-\sqrt{1/2}-O(\epsilon^{1/3})$.

\paragraph{Least Median of Squares (LMS).}
These methods~\cite{rousseeuw-robust-87} in robust statistics seek a
fit that minimizes the median residual value separating the fit from
the sample points.  This is not a depth-based criterion, but it leads
to fits which are highly robust against outliers.  For location
problems, the least median of squares fit is the center of the minimum
radius sphere that contains at least half of the sample
data~\cite{harpeled-kdisc-03}.  It has a breakdown point of $\frac12$:
if fewer than half the sample data points are outliers, then the sphere
defining the LMS fit has smaller radius than the circumsphere of the
non-outliers, and it contains at least one non-outlier, so its center
must be an accurate fit.  Clearly, this is the best breakdown point
possible for any location method.  The natural type of
$\epsilon$-approximation to use for this problem is one with balls as
its ranges.  If we form the LMS fit of such an
$\epsilon$-approximation, the result may not be robust.  Instead, we
approximate the LMS fit by finding the center of the minimum radius
sphere that contains at least a $\frac12+\epsilon$ proportion of the
points in the $\epsilon$-approximation.  Such a sphere must therefore
contain at least half of the sample data, and has a radius at least as
small as the smallest sphere containing  at least a $\frac12+2\epsilon$
fraction of the sample data.  It is robust with a breakdown point of
$\frac12-2\epsilon$.

The same LMS approach can also be applied to
regression problems.  The least median of squares regression hyperplane
can be defined as the central hyperplane in a slab bounded by two
parallel hyperplanes, with minimum vertical separation between them,
that contains at least half of the sample data; again this is robust
with a breakdown point of $\frac12$.  As above, we can use an
$\epsilon$-approximation, with slab ranges, and find the slab with
minimum vertical separation containing a $\frac12+\epsilon$ fraction of
the $\epsilon$-approximation points, to produce an approximate LMS fit
with breakdown point $\frac12-2\epsilon$.

%=====================================================================
%  Lower Bounds
%=====================================================================

\section{A Lower Bound on Range Counting} 
\label{sec:lowerbounds}

We provide a simple lower bound on the space required to count
approximately the number of items in a range that is not necessarily
an iceberg. When we say that an
algorithm $f$-approximates the range counting problem we mean that if
a given range contains $l$ points, the algorithm gives us an answer
which lies between $l/f$ and $l\cdot f$.

The bound is stated in terms of {\em two-sided ranges}: a point
$(x,y)$ is said to belong to the two sided range located at $(p,q)$ if
$x \geq p$ and $y \geq q$.

\begin{theorem}
\label{thm:lb_rangecounting}
Any $f$-approximate algorithm to the two-sided range counting problem
must use space $\Omega(n/f^2)$. 
\end{theorem}

We begin by assuming there is an algorithm $A$ which gives an $f$
approximation to the two-sided range counting problem for a stream of
points in two dimensions. Further we assume that this algorithm uses
space $o(n/f^2)$.

Now consider a set of $n$ points which are grouped in $n/f^2$ equally
sized groups, we call them $G_i$, where $1 \leq i \leq n/f^2$, in the
following way:

\begin{itemize}
\item Each point in $G_i$ has the same $x$ coordinate, we call it
  $x_i$. Additionally, we require $x_i > x_{i-1}$.
\item All the points in $G_i$ have $y$ coordinates closely clustered
  at a given value, we call it $y_i$. Formally, for every $p_j \in
  G_i$, we say that $0 \leq y(p_j) - y_i < 1/2$.
\item Every point $p_j \in G_i$ has $y$-coordinate strictly smaller
  than the $y$-coordinates of all the points in $G_{i-1}$.
\item Each group $i$ has an additional point $q_i = (x_i+\epsilon,
  y_i+\epsilon)$, for some $\epsilon < 1/2$, associated with it.
\end{itemize}

Note that this family of input sequences has the property that a
two-sided query made at $(x_i,y_i)$ should return a count of $f^2 + 1$
and one made at $(x_i+\frac{\epsilon}{2},y_i+\frac{\epsilon}{2})$
should return a count of 1. This radical change in the counts will not
occur between two such queries at any point which is not actually
$(x_i, y_i)$ for some value of $i$. As an extension to this simple
observation, we note that if all the $x_i$s and $y_i$s are chosen out
of the integers $1, 2, \ldots n$, it is possible to extract the exact
values of all the $x_i$ with exactly $2n^2$ queries. 

Let us see if the algorithm $A$ can be the query mechanism which we
can deploy to this end. Since $A$ is an $f$ approximation, it should
return a value of at most $f$ at
$(x_i+\frac{\epsilon}{2},y_i+\frac{\epsilon}{2})$ and a value between
$f + 1/f$ and $f^3 + f$ at $(x_i,y_i)$. This means that $A$ can indeed
act as the oracle which identifies the locations of the groups in our
set.

Hence, using $A$ as a subroutine we can extract $\theta(n/f^2)$
information about the input set. This contradicts the assumption that
$A$ uses space $o(n/f^2)$. \qed

Seen in the context of streaming algorithms,
Theorem~\ref{thm:lb_rangecounting} implies that is not possible to
approximate the range counting problem in polylogarithmic space. One
of the implications of this, among others, is that it is not possible
to count inversions in lists~\cite{gupta-inversions-03} in the sliding
window model.

{\small \paragraph{Acknowledgments.} We would like to thank David Mount
for helpful discussions of robust statistics in the context of the
topics of this paper, and S. Muthukrishnan for helpful discussions on
geometric streaming algorithms in general.}

{\footnotesize\raggedright
\bibliographystyle{abbrv}
\bibliography{streams,geom}
}

%\appendix

\end{document}